\documentclass[a4paper, amsfonts, amssymb, amsmath, reprint, showkeys, nofootinbib, twoside]{revtex4-1}
\usepackage[english]{babel}
\usepackage[utf8]{inputenc}
\usepackage[colorinlistoftodos, color=green!40, prependcaption]{todonotes}
\usepackage{amsthm}
\usepackage{mathtools}
\usepackage{physics}
\usepackage{xcolor}
\usepackage{graphicx}
\usepackage[left=23mm,right=13mm,top=35mm,columnsep=15pt]{geometry} 
\usepackage{adjustbox}
\usepackage{placeins}
\usepackage[T1]{fontenc}
\usepackage{lipsum}
\usepackage{csquotes}

\usepackage{amssymb}
\usepackage{natbib} 
\usepackage{comment}

\usepackage[LGRgreek]{mathastext} 
\newcommand\tab[1][1cm]{\hspace*{#1}} 

\usepackage{lineno}
\usepackage{setspace}
\usepackage{hyperref}

\begin{document}
\title{Enhanced photocatalytic dye degradation and hydrogen production ability of Bi\textsubscript{25}FeO\textsubscript{40}-rGO nanocomposite and mechanism insight}

\author{M. A. Basith\textsuperscript{*}\email[Email address: ]{mabasith@phy.buet.ac.bd}, Ragib Ahsan, Ishrat Zarin, and M. A. Jalil}
    
    \affiliation{Nanotechnology Research Laboratory, Department of Physics, Bangladesh University of Engineering and Technology, Dhaka-1000, Bangladesh.\\
    \textsuperscript{*}Corresponding author: mabasith@phy.buet.ac.bd\\
    \\ DOI: \href{https://www.nature.com/articles/s41598-018-29402-w}{10.1038/s41598-018-29402-w}}


\begin{abstract}
A comprehensive comparison between Bi$_{25}$FeO$_{40}$-reduced graphene oxide(rGO) nanocomposite and BiFeO$_3$-rGO nanocomposite has been performed to investigate their photocatalytic abilities in degradation of Rhodamine B dye and generation of hydrogen by water-splitting. The hydrothermal technique adapted for synthesis of the nanocomposites provides a versatile temperature-controlled phase selection between perovskite BiFeO$_3$ and sillenite Bi$_{25}$FeO$_{40}$. Both perovskite and sillenite structured nanocomposites are stable and exhibit considerably higher photocatalytic ability over pure BiFeO$_3$ nanoparticles and commercially available Degussa P25 titania. Notably, Bi$_{25}$FeO$_{40}$-rGO nanocomposite has demonstrated superior photocatalytic ability and stability under visible light irradiation than that of BiFeO$_3$-rGO nanocomposite. The possible mechanism behind the superior photocatalytic performance of Bi$_{25}$FeO$_{40}$-rGO nanocomposite has been critically discussed.
\end{abstract}


\maketitle

\section{Introduction}

High expenses associated with fabrication of complex optoelectronic devices, coupled with their poor conversion efficiency, limits the use of abundant solar energy from the sun \cite{ref1,ref2,ref3,ref4,ref5}. Besides overcoming these obstacles, the researchers have sought to convert solar energy into a more usable and reproducible form which is also inexpensive to store. Interestingly, substrateless photocatalyst powders suspended in earth-abundant water can absorb the sunlight incident on them and electrochemically split the water molecules to produce hydrogen and oxygen \cite{refrev2,refrev14}. Storing the generated hydrogen, a green fuel, is also easy and inexpensive. Furthermore, these photocatalyst powders can be used to degrade different industrial pollutants which are excessively harmful to our environment \cite{refrev1,refrev3,refrev4,refrev5,refrev6,refrev7,refrev8}. Since photocatalysis has been able to manifest a great potential in the production of hydrogen by splitting water molecules and degradation of environmental pollutants, researchers have been thoroughly investigating it for both energy-related and environmental applications \cite{refrev6,refrev7,ref7,ref9,ref11}. Metal oxide photocatalysts are being studied extensively for their enhanced catalytic abilities supplemented by highly stable nature under the electrochemical reaction conditions \cite{ref9,ref11}. An efficient photocatalyst is expected to have a large surface area, superior sensitivity to the visible region of the solar spectrum, appropriate band energetics, and agile carrier transport to inhibit recombination processes\cite{ref7,ref11,ref8,ref12}. While graphene is not well-suited for optical applications due to the absence of band gap, its enormous surface area and extraordinary transport properties can be used to supplement the photocatalytic abilities of metal oxides \cite{refrev5,refrev8,refX,refY,refZ,refrev9,refrev10,refrev11}. Investigations on the in-situ growth of metal oxide nanoparticles on the 2D lattice of graphene have unveiled incredible improvements in morphology, optical sensitivity, adsorption capabilities, and electron transport characteristics over the pristine metal oxide nanoparticles \cite{refrev9,refrev10,refrev11,refXX,refYY,refZZ}. Furthermore, several investigations on engineering the work function of graphene implied that it is capable of modifying the energy bands of a semiconductor by forming heterojunction \cite{refrev13,ref500,ref501,ref502}.

Among numerous metal oxides, the ones belonging to perovskite and sillenite structures have drawn more interest as photocatalysts as they have shown good optical response and great stability \cite{ref9,ref11,refXXX,refYYY,refRSC}. Bismuth ferrite (BFO) is one such family of metal oxide that shows both perovskite (BiFeO$_3$) and sillenite (Bi$_{25}$FeO$_{40}$) structures \cite{refZZZ,refXXXX}. Both BiFeO$_3$ and Bi$_{25}$FeO$_{40}$ have reasonably narrow band gaps of 2.1$\sim$2.7 eV and 1.8$\sim$2.0 eV respectively as reported in previous investigations \cite{ref42,refYYYY,refZZZZ,refscirep2}. While Bi$_{25}$FeO$_{40}$ has a better optical absorption capability due to the smaller band gap, ferroelectric nature of BiFeO$_3$ helps it develop a superior resistance to recombination processes \cite{ref42,ref1718}. However, until now, it has been challenging to synthesize both BiFeO$_3$ and Bi$_{25}$FeO$_{40}$ nanoparticles with excellent phase-purity due to the difficulties associated with controlling the synthesis parameters, in particular, synthesis temperature \cite{ref1718,ref1719,ref1720,ref1721}. Moreover, unfavorable band energetics of both BiFeO$_3$ and Bi$_{25}$FeO$_{40}$ have also limited their use in energy-related applications, e.g., photocatalytic hydrogen production from water. Notably, graphene lattice may work as a center for heterogeneous nucleation to favor single-phase growth of either perovskite or sillenite structure depending on the synthesis conditions \cite{ref1722,ref1723,ref1724}. The formation of a heterojunction with graphene may improve the photocatalytic hydrogen production ability of both BiFeO$_3$ and Bi$_{25}$FeO$_{40}$ by reducing their work functions and modifying the band energetics. Inspired by the possibility of such improvements, we have incorporated reduced graphene oxide (rGO) with both BiFeO$_3$ and Bi$_{25}$FeO$_{40}$ to fabricate their nanocomposites. We have demonstrated a hydrothermal synthesis process for BFO-rGO nanocomposites that can be utilized to obtain both perovskite (BiFeO$_3$-rGO) and sillenite (Bi$_{25}$FeO$_{40}$-rGO) structures by tuning the hydrothermal reaction temperature \cite{ref42}. In this present investigation, we have compared the performance of BiFeO$_3$-rGO and Bi$_{25}$FeO$_{40}$-rGO nanocomposites to that of both pristine BiFeO$_3$ nanoparticles and commercially available Degussa P25 titania nanoparticles in photocatalytic degradation of Rhodamine B (RhB) dye and photocatalytic hydrogen production via water-splitting. Here, RhB dye has been used as a proxy to the industrial pollutants to benchmark the photocatalytic ability of the aforementioned photocatalysts in environmental applications. Through a series of experiments, we have observed that Bi$_{25}$FeO$_{40}$-rGO nanocomposite exhibits a considerably better visible light driven photocatalytic performance in both dye degradation and hydrogen production than BiFeO$_3$-rGO nanocomposite, BiFeO$_3$ nanoparticles, and Degussa P25 nanoparticles. Furthermore, we have critically assessed the possible mechanism behind the superior photocatalytic performance of Bi$_{25}$FeO$_{40}$-rGO nanocomposite.

\section{Methods}

\subsection{Synthesis of BiFeO$_3$ nanoparticles, BiFeO$_3$-rGO, and Bi$_{25}$FeO$_{40}$-rGO nanocomposites}

Synthesis of BiFeO$_3$ nanoparticles, BiFeO$_3$-rGO, and Bi$_{25}$FeO$_{40}$-rGO nanocomposites was carried out using the hydrothermal technique. Bi(NO$_3$)$_3$.5H$_2$O (Sigma-Aldrich, 98.0\%) and Fe(NO$_3$)$_3$.9H$_2$O (Sigma-Aldrich, 98.0\%) were used as precursors during synthesis of pure BiFeO$_3$ nanoparticles. Graphene oxide (GO) synthesized using modified Hummer's method was used as an additional precursor for the synthesis of the desired nanocomposites. Pure BiFeO$_3$ nanoparticles were synthesized at a hydrothermal reaction temperature of 180 $^o$C. Most interestingly, a synthesis temperature of 180 $^o$C yielded the Bi$_{25}$FeO$_{40}$-rGO nanocomposite whereas the BiFeO$_3$-rGO nanocomposite required an elevated temperature of 200 $^o$C. The synthesis procedure and effects of hydrothermal reaction temperature on the crystal structure of the nanocomposites have been elucidated in our previous investigation \cite{ref42}.

\subsection{Characterizations Techniques}

Structural characterization of the prepared samples was performed by obtaining their X-ray diffraction (XRD) patterns using a diffractometer (PANalytical Empyrean) with a Cu X-ray source (Wavelength, ${\lambda}$: K$_{\alpha1}$= 1.540598 \AA $ $ and K$_{\alpha2}$ = 1.544426 \AA). Field emission scanning electron microscopy (FESEM) imaging of the samples was performed using a scanning electron microscope (XL30SFEG; Philips, Netherlands). The FESEM images were analyzed to determine the morphology of the nanomaterials. ImageJ software was used to analyze the size of the particles. X-ray photoelectron spectroscopy (XPS) was used to investigate the surface chemical states of synthesized nanomaterials. Fourier transform infrared (FTIR) spectroscopy (Shimadzu, IRSpirit-T) was also performed in the wavenumber range of 1000 cm$^{-1}$ to 4000 cm$^{-1}$ for studying the presence of different functional groups. Steady-state photoluminescence (PL) spectroscopy (Hitachi, F-700 FL spectrophotometer) was used to investigate the photogenerated carrier recombination of the nanomaterials. The excitation wavelength used for PL spectroscopy was 400 nm.

\subsection{Photocatalytic degradation test} 
A photocatalytic degradation test was performed on Rhodamine B(RhB) dye using the synthesized nanomaterials as photocatalyst following the typical procedures. A solution was prepared by dissolving 10 mg RhB in 50 ml of distilled water. An aliquot of 5 ml was extracted from the solution to measure its absorbance using a UV-vis spectrophotometer. The intensity of the absorbance peak (553 nm) can be considered proportional to the amount of RhB present in the solution. This method was used each time when it was required to determine the remnant amount of RhB in the solution. Then the solution was magnetically stirred in dark conditions for 30 minutes to obtain a homogeneous solution. 40 mg of photocatalyst was then added to the solution and again stirred for 30 minutes to ensure adsorption-desorption equilibrium between photocatalyst and RhB. The photocatalysis process was initiated by irradiating the suspension with a Mercury-Xenon lamp (Hamamatsu L8288, 500W). The UV portion of the light was filtered out with a UV cut-off filter (\textless 400 nm). The degradation of RhB dye was monitored by carrying out absorbance measurements of the suspension at every 30 minutes interval. However, the photocatalyst nanomaterials were precipitated from the suspension using a centrifuge machine at 5500 rpm for 10 minutes before measuring its absorbance. Prior to recycling the photocatalyst, the remnant photocatalyst nanomaterials in the suspension were separated by centrifugation, washed with distilled water and dried afterward.

\subsection{Photocatalytic hydrogen generation test}
Photocatalytic hydrogen production test was performed in a slurry-type photochemical reactor. 20 mg of photocatalyst sample was dispersed into 30 ml deionized water, and the suspension was stirred magnetically to obtain homogeneity. The inert atmosphere required for photocatalytic hydrogen production was ensured by purging the system with argon gas for 30 minutes. The suspension was irradiated with visible light by a Mercury-Xenon lamp (Hamamatsu L8288, 500W) coupled with a UV cut-off filter (\textless 400 nm). The generated gas was extracted at an interval of 30 minutes for 4 hours and evaluated by a gas chromatography (GC) device equipped with a thermal conductivity detector (TCD) and a gas analyzer. The amount of generated hydrogen gas was determined in ml H$_2$/g catalyst.

\section{Results and Discussions}

\subsection{Materials characterization}

The crystallographic structure of the synthesized nanomaterials has been investigated by obtaining their X-ray diffraction (XRD) patterns as shown in Fig. 1. The diffraction peaks exhibited by both BiFeO$_3$ and BiFeO$_3$-rGO conform to the single-phase perovskite structure (R3c space group) of BiFeO$_3$ (JCPDS card No. 86-1518). On the other hand, the diffraction peaks of Bi$_{25}$FeO$_{40}$-rGO are in good agreement with single-phase sillenite structure (I23 space group) of Bi$_{25}$FeO$_{40}$ (JCPDS card No. 46-0416). Reduction of GO can be observed for both nanocomposites as the characteristic peak of GO at (002) is absent in the corresponding patterns. The average particle size of the synthesized nanomaterials has been obtained from particle size histograms extracted from their field emission scanning electron microscopy (FESEM) images. Incorporation of rGO has reduced the average particle size of BiFeO$_3$-rGO (40 nm) and Bi$_{25}$FeO$_{40}$-rGO (20 nm) nanocomposites compared to that of pure BiFeO$_3$ nanoparticles (70 nm) as shown in supplemental Fig. S1. Comparatively smaller particle size implies larger surface to volume ratio which is a necessary property of an efficient photocatalyst. The mechanism behind inhibition of particle growth due to the inclusion of rGO has been presented elsewhere \cite{ref42}. 

\begin{figure}
  \centering
  \includegraphics[width= 0.5\textwidth]{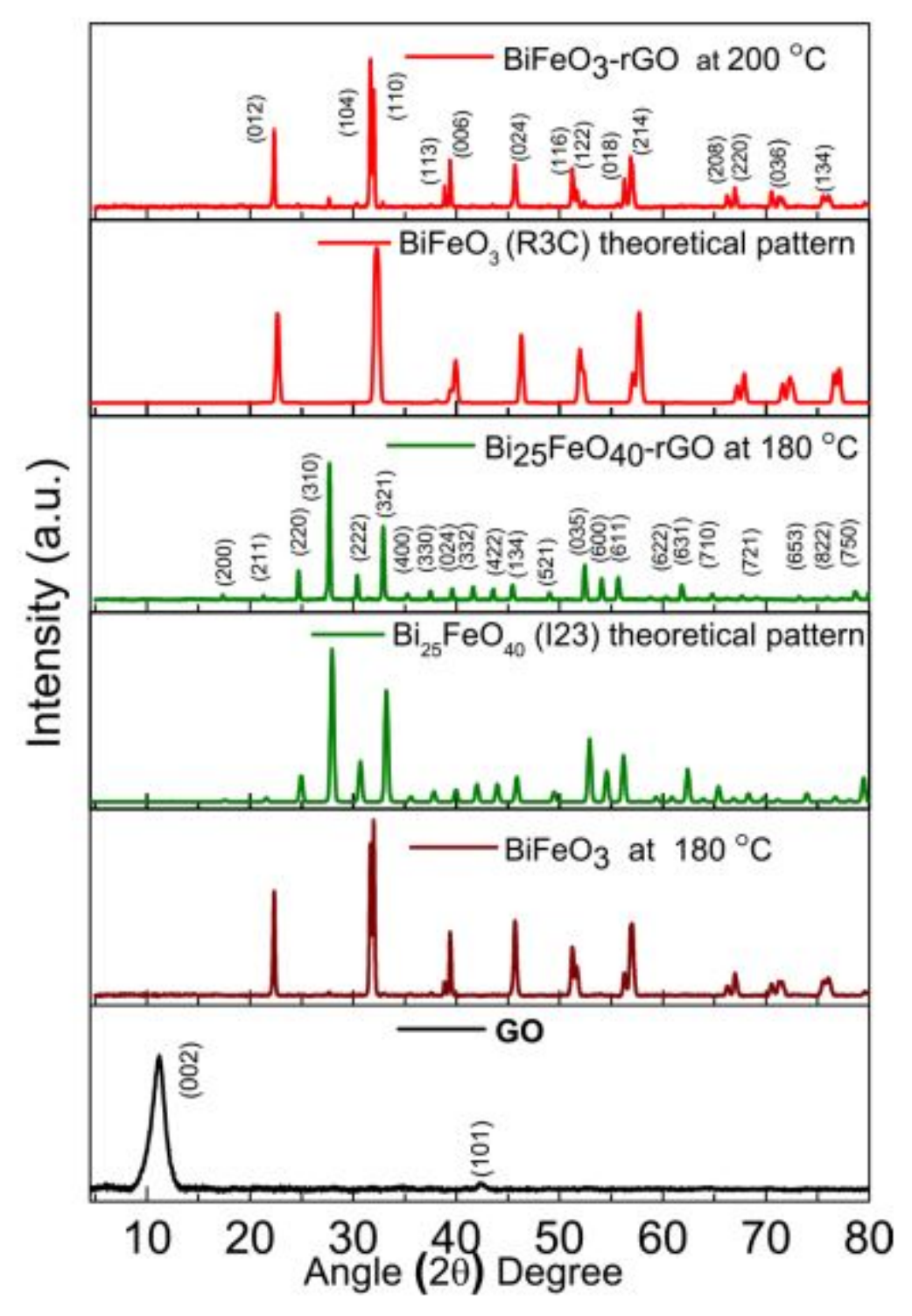}
  \caption{XRD patterns of GO, BiFeO$_3$ at 180 $^o$C, sillenite type Bi$_{25}$FeO$_{40}$-rGO at 180 $^o$C,  perovskite type BiFeO$_3$-rGO nanocomposite at 200 $^o$C. The theoretically calculated patterns for BiFeO$_3$ of space group R3C and Bi$_{25}$FeO$_{40}$ of space group \textit{I23} has also been included for comparison of phases.}
\end{figure}

\tab X-ray photoelectron spectroscopy (XPS) has been performed on the synthesized nanomaterials to analyze the formation of different bonds on their surface. Fig. 2(a-c) presents the O 1s spectra of BiFeO$_3$ nanoparticles, BiFeO$_3$-rGO, and Bi$_{25}$FeO$_{40}$-rGO nanocomposites. The peaks for both Bi-O and Fe-O bonds occur at around 529 eV and 530 eV respectively for all the samples. This result is consistent with the values reported in the literature \cite{refXPSper,refXPSsil,refscirep1}. As shown in Fig. 2(b-c), the presence of a peak around 531 eV can be attributed to the formation of COOR bond, i.e., the formation of a composite with graphene \cite{refXY}. Fig. 2(d-f) shows the C 1s spectra of Bi$_{25}$FeO$_{40}$-rGO, BiFeO$_3$-rGO nanocomposites, and GO respectively. As shown in Fig. 2(f), the C-O peak is almost as strong as the C-C peak in GO. However, the intensity of the C-O peak is significantly reduced in both nanocomposites as shown in Fig. 2(d-e) while the intensity of the COOR peak does not change significantly. The dominance of the C-C peak compared to the C=O and COOR peaks implies that GO has been considerably reduced and the honeycomb structure of graphene has also been nearly restored. This partial reduction of GO can be attributed to the nature of reduction, e.g., the hydrothermal process. The hydrothermal process reduces the GO sheets by decomposing the C=O and COOR bonds \cite{refhydro1}. Higher thermal energy provided by a higher process temperature favors faster and greater reduction \cite{refhydro1,refhydro2,refhydro3,refhydro4,refhydro5,refhydro6,refhydro7}. However, a process temperature that is higher than 180 $^o$C would be unfavorable for the single phase growth of Bi$_{25}$FeO$_{40}$-rGO nanocomposite \cite{ref42}. Although single phase BiFeO$_3$-rGO nanocomposite can be synthesized at a higher temperature, the higher growth rate may cause the particles to grow larger and even agglomerate. In addition to the higher expenses associated with a higher temperature process, we may face further difficulty in the synthesis process as the control over the particle size and morphology might be lost. Besides, the process also removes carbon atoms from the carbon plane in the form of CO$_2$ and splits the graphene sheet into smaller pieces \cite{refhydro7,refhydro8}. The dangling carbon atoms at the end of the smaller graphene sheets may be oxidized again and contribute to the COOR bonds present in the surface of the composites.

\begin{figure*}
 \centering
 \includegraphics[width=  1\textwidth]{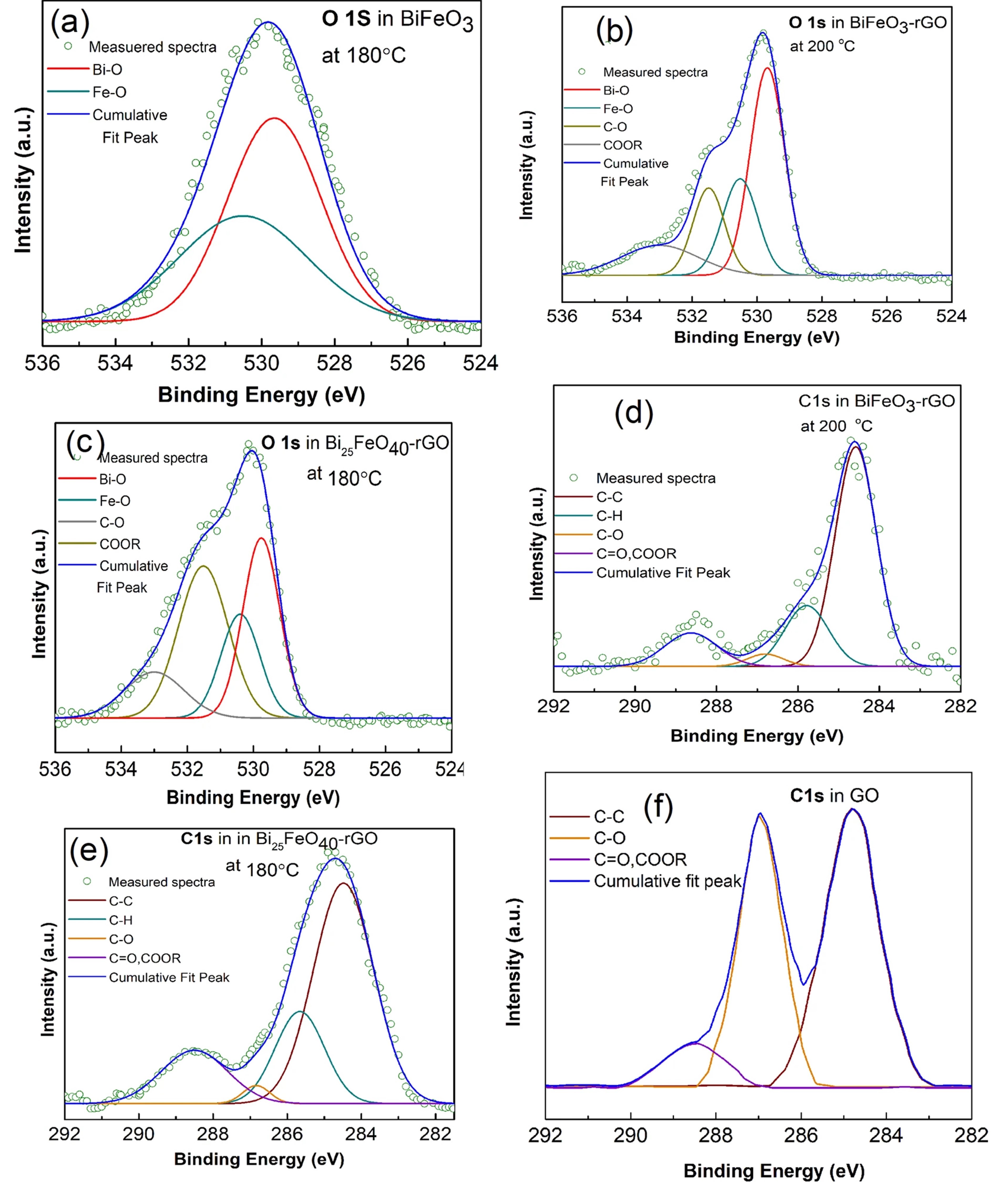}
 \caption{ XPS spectra of the core level O 1s of (a) BiFeO\textsubscript{3}, (b) perovskite BiFeO\textsubscript{3}-rGO nanocomposite, (c) sillenite Bi\textsubscript{25}FeO\textsubscript{40}-rGO nanocomposite, and core level C 1s of (d) perovskite BiFeO\textsubscript{3}-rGO nanocomposite, (e) sillenite Bi\textsubscript{25}FeO\textsubscript{40}-rGO nanocomposite, and (f) GO.} 
\end{figure*}

\tab Since XPS analyzes the bond properties of the surface (within 3-5 nm) of a sample rather than its bulk, an analysis like FTIR spectroscopy would be more useful for studying the presence of different functional groups in the nanocomposites. Hence, FTIR spectroscopy has also been performed on GO and the nanocomposites to verify the reduction of GO into rGO conclusively. The components and functional groups have been identified from the FTIR spectra as shown in Fig. 3. The vibrational band ranging from 1053 to 1220 cm$^{-1}$ can be attributed to the presence of different types of C-O bonds. In addition, the vibrational bands at 1718 cm$^{-1}$ and 3441 cm$^{-1}$ can be assigned to the presence of C=O and -OH bonds respectively \cite{refftir1,refftir2,refftir3,refftir4,refftir5}. As evident from Fig. 3, the peaks for each of these oxidized species are considerably smaller in the spectra of the nanocomposites compared to that of GO. Notably, the peak due to C=O and COOR bonds is significantly smaller than that of GO for both nanocomposites. Hence, the FTIR spectra prove that even the C=O and COOR bonds have also been considerably reduced. FTIR characterization, in conjunction with the XPS characterization, leads us to the conclusion that GO has been reduced to a significant extent during the formation of the nanocomposites.

\begin{figure}
  \centering
  \includegraphics[width= 0.5\textwidth]{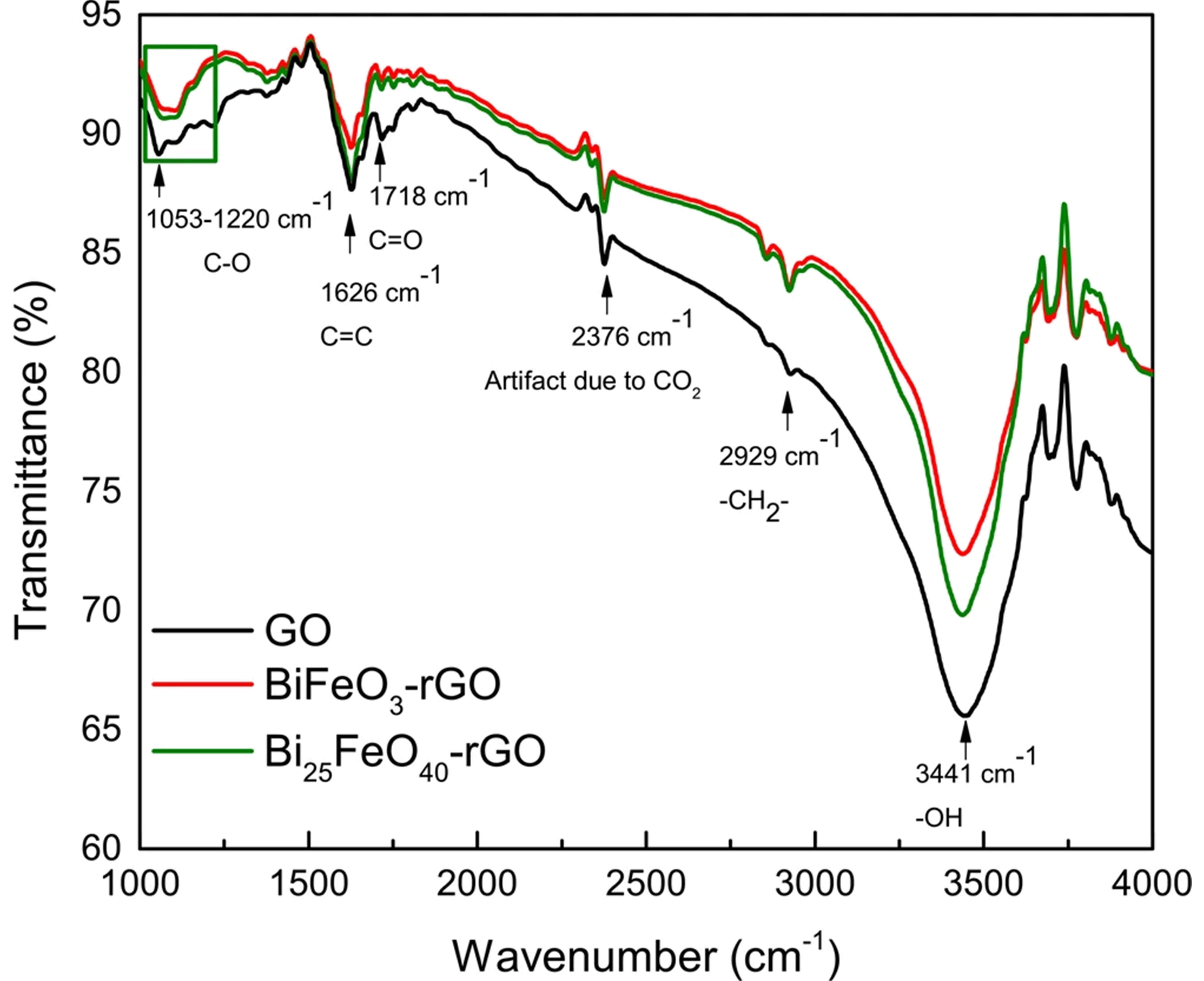}
  \caption{FTIR spectra of GO, BiFeO\textsubscript{3}-rGO nanocomposite, and Bi\textsubscript{25}FeO\textsubscript{40}-rGO nanocomposite.}
\end{figure}

\subsection{Photocatalytic Degradation Activity and Stability}

BiFeO$_3$ nanoparticles, BiFeO$_3$-rGO, and Bi$_{25}$FeO$_{40}$-rGO nanocomposites have been used as the photocatalysts to perform degradation of RhB dye under visible light irradiation. For comparison, Bi$_{25}$FeO$_{40}$ nanoparticles and commercially available Degussa P25 titania nanoparticles were also used to perform the RhB dye degradation experiment under the same experimental conditions. Fig. 4(a) presents the time-dependent UV-vis absorbance spectra of RhB when Bi$_{25}$FeO$_{40}$-rGO is used as the photocatalyst. Fig. 4(b) shows the photocatalytic efficiency of BiFeO$_3$ nanoparticles, Bi$_{25}$FeO$_{40}$ nanoparticles, P25 titania nanoparticles, BiFeO$_3$-rGO and Bi$_{25}$FeO$_{40}$-rGO nanocomposites in the degradation of RhB. As observed in Fig. 4(b), BiFeO$_3$ nanoparticles, Bi$_{25}$FeO$_{40}$ nanoparticles, BiFeO$_3$-rGO and Bi$_{25}$FeO$_{40}$-rGO nanocomposites can degrade 60\%, 62\%, 78\%, and 87\% of the RhB dye respectively after 4 hours of irradiation compared to the 6\% degradation achieved by commercially available Degussa P25 titania nanoparticles. The degradation rates of the photocatalysts have also been calculated from Fig. 4(c) using the Langmuir-Hinshelwood model. The degradation rates are found to be 6.4 $\times$ 10$^{-4}$ min$^{-1}$, 2.83 min$^{-1}$, 2.88 min$^{-1}$, 4.45 min$^{-1}$, and 5.91 min$^{-1}$ for Degussa P25, BiFeO$_3$, Bi$_{25}$FeO$_{40}$, BiFeO$_3$-rGO, and Bi$_{25}$FeO$_{40}$-rGO respectively. Notably, Bi$_{25}$FeO$_{40}$-rGO displays the highest photocatalytic degradation ability as it exhibits a 108\%, 104\%, and 33\% greater degradation rate compared to that of BiFeO$_3$, Bi$_{25}$FeO$_{40}$, and BiFeO$_3$-rGO respectively.

\begin{figure*}
  \centering
  \includegraphics[width = 1\textwidth]{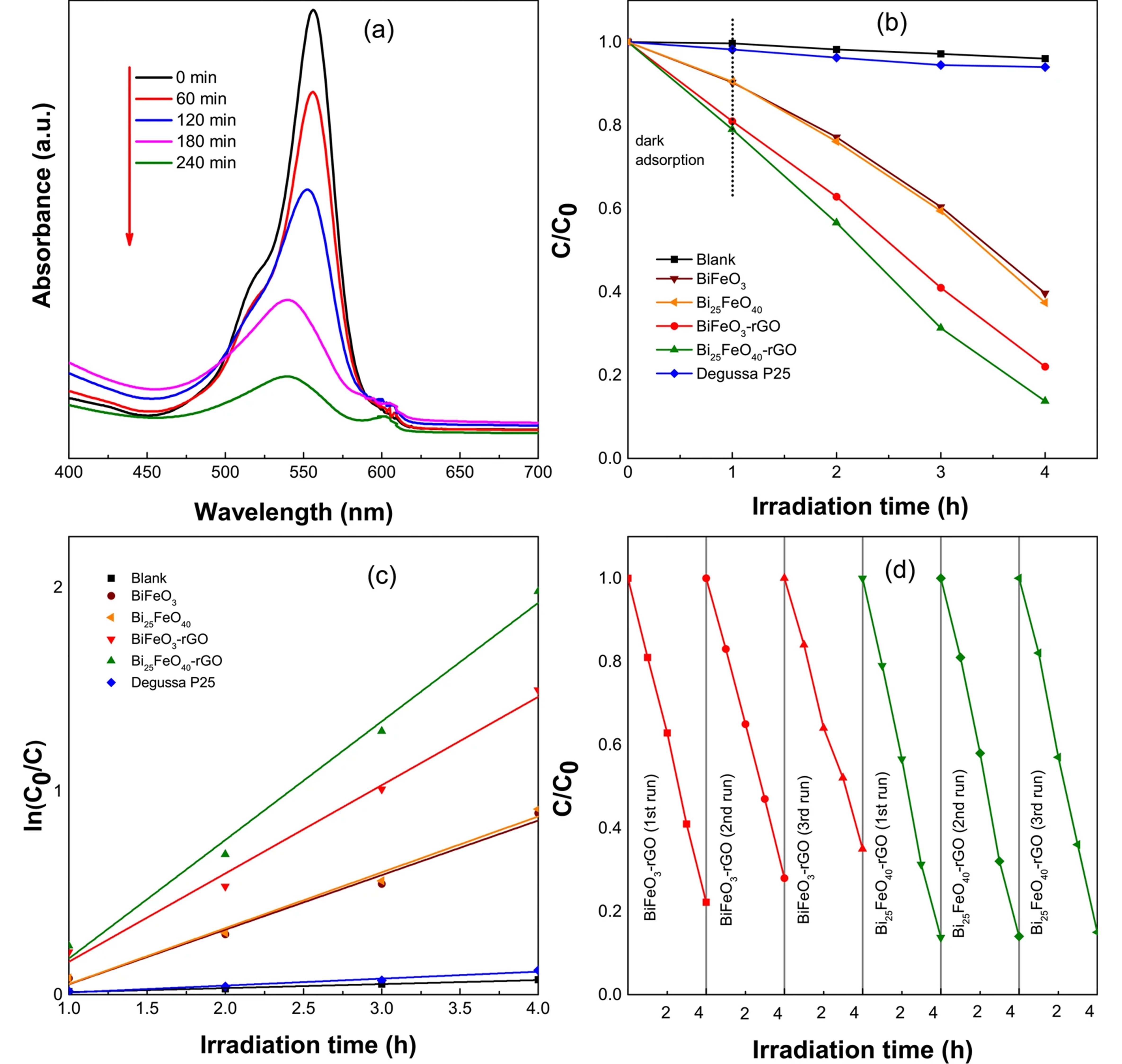}
  \caption{ (a) Time-dependent absorbance spectra of RhB solution under visible light irradiation with Bi\textsubscript{25}FeO\textsubscript{40}-rGO nanocomposite photocatalyst, (b) Degradation of RhB by photocatalytic activity of blank, BiFeO\textsubscript{3}, Bi\textsubscript{25}FeO\textsubscript{40}, BiFeO\textsubscript{3}-rGO, Bi\textsubscript{25}FeO\textsubscript{40}-rGO, and Degussa P25 with irradiation time, (c) Pseudo-first order kinetics fitting data for RhB photodegradation by blank, BiFeO\textsubscript{3}, Bi\textsubscript{25}FeO\textsubscript{40}, BiFeO\textsubscript{3}-rGO, Bi\textsubscript{25}FeO\textsubscript{40}-rGO, and Degussa P25 (d) Recyclability of BiFeO\textsubscript{3}-rGO and Bi\textsubscript{25}FeO\textsubscript{40}-rGO for 3 successive runs.}
\end{figure*}

\tab In addition to the superior degradation ability, a photocatalyst is expected to be stable under the reaction conditions to be considered for practical applications. Since the stability of BiFeO$_3$ as a photocatalyst is well known, only BiFeO$_3$-rGO and Bi$_{25}$FeO$_{40}$-rGO nanocomposites were subject to the recyclability test. As observed in Fig. 4(d), BiFeO$_3$-rGO exhibits a small loss in degradation ability during each successive cycle. However, Bi$_{25}$FeO$_{40}$-rGO shows excellent stability with negligible loss in photocatalytic ability after 3 consecutive cycles. Since Bi$_{25}$FeO$_{40}$-rGO nanocomposite has shown a comparatively higher photocatalytic degradation ability and better stability than BiFeO$_3$-rGO nanocomposite, we believe that it can be considered for different practical applications such as environmental pollutant degradation, hydrogen production via water-splitting, etc.

\subsection{Photocatalytic Hydrogen Production Test Analysis}
A photocatalytic hydrogen production experiment has also been performed using the synthesized nanomaterials. Fig. 5 presents the amount of hydrogen gas (ml/g) generated after 4 hours of visible light irradiation. Photocatalytic hydrogen production of BiFeO$_3$, BiFeO$_3$-rGO, and Bi$_{25}$FeO$_{40}$-rGO has also been compared to that of a standard photocatalyst, Degussa P25 titania nanoparticles \cite{refp25}. Fig. 5 presents the amount of hydrogen produced by BiFeO$_3$, BiFeO$_3$-rGO, Bi$_{25}$FeO$_{40}$-rGO and Degussa P25 under the same experimental conditions. In addition to the high photocatalytic degradation ability, Bi$_{25}$FeO$_{40}$-rGO nanocomposite demonstrates the highest volume of hydrogen production which is almost 3.5 times greater than that of Degussa P25 nanoparticles.

\begin{figure}
  \centering
  \includegraphics[width= 0.5\textwidth]{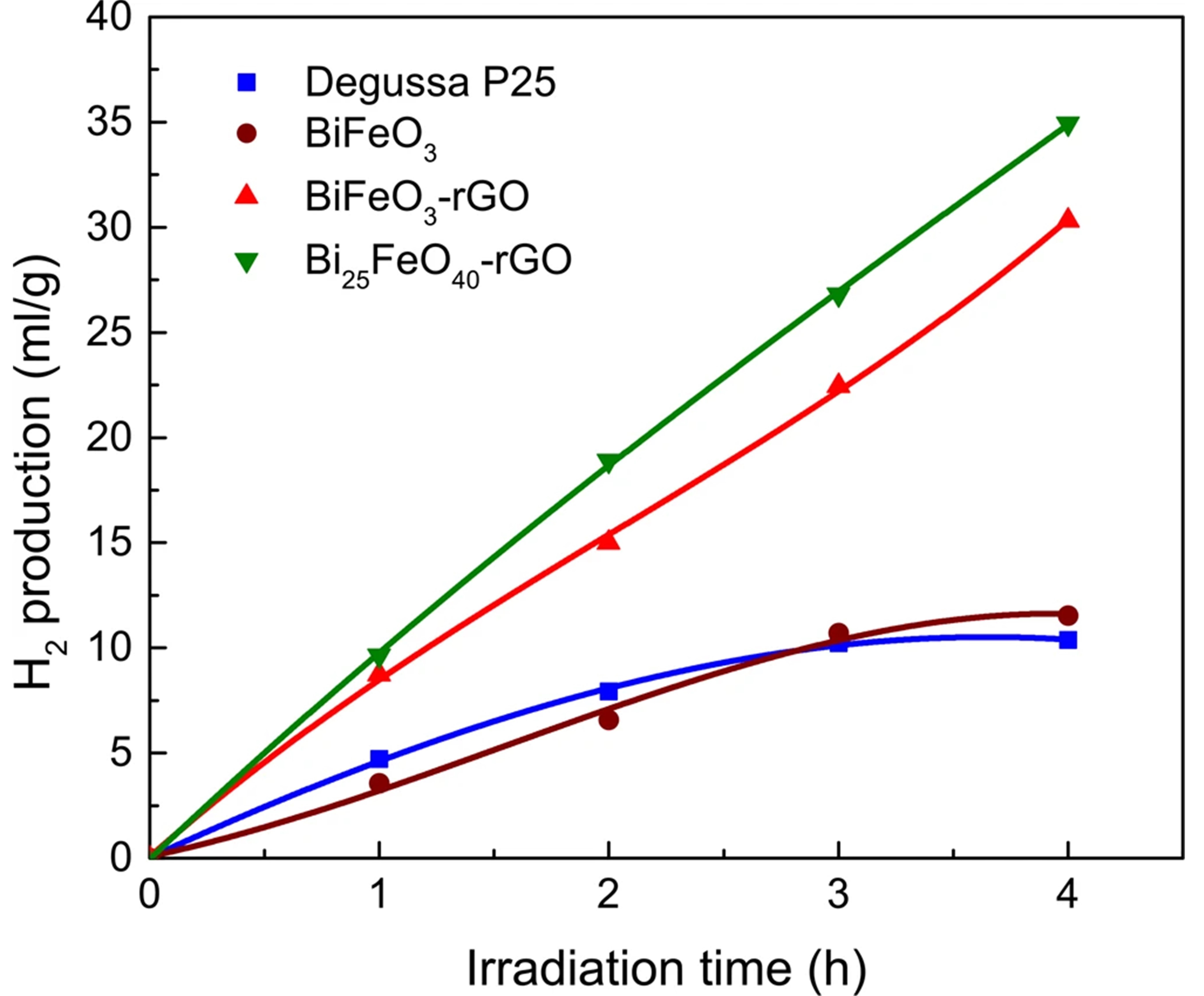}
  \caption{ The volume of evolved hydrogen versus irradiation time for Degussa P25, BiFeO\textsubscript{3}, BiFeO\textsubscript{3}-rGO, and Bi\textsubscript{25}FeO\textsubscript{40}-rGO.}
\end{figure}

\subsection{Comparison of Optical Band gap}
The photocatalytic behavior of the prepared catalysts in both dye degradation and hydrogen production can be rationalized by analyzing their optical absorption, band edge positions, and possible mechanisms to suppress the recombination processes. The optical band gaps of BiFeO$_3$, Bi$_{25}$FeO$_{40}$, BiFeO$_3$-rGO and Bi$_{25}$FeO$_{40}$-rGO are found to be 2.1 eV, 1.8 eV, 1.8 eV and 1.65 eV respectively. The UV-vis absorbance spectra of these materials have been provided in the supplemental Fig. S2 \cite{ref41,ref42}. Hence, with a smaller band gap, Bi$_{25}$FeO$_{40}$-rGO nanocomposite possesses greater optical absorption ability compared to that of BiFeO$_3$-rGO nanocomposite and BiFeO$_3$ nanoparticles. In addition to absorbing photons belonging to the infrared region of the solar spectrum, the band gap of 1.65 eV enables the Bi$_{25}$FeO$_{40}$-rGO nanocomposite in nearly reaching the Shockley-Queisser limit which states that maximum solar energy conversion can be achieved optoelectronically with a material that has an optical band gap of 1.34 eV \cite{refABCD,refABDC,refACBD,refzyz}.

\begin{figure*}
  \centering
  \includegraphics[width = 1\textwidth]{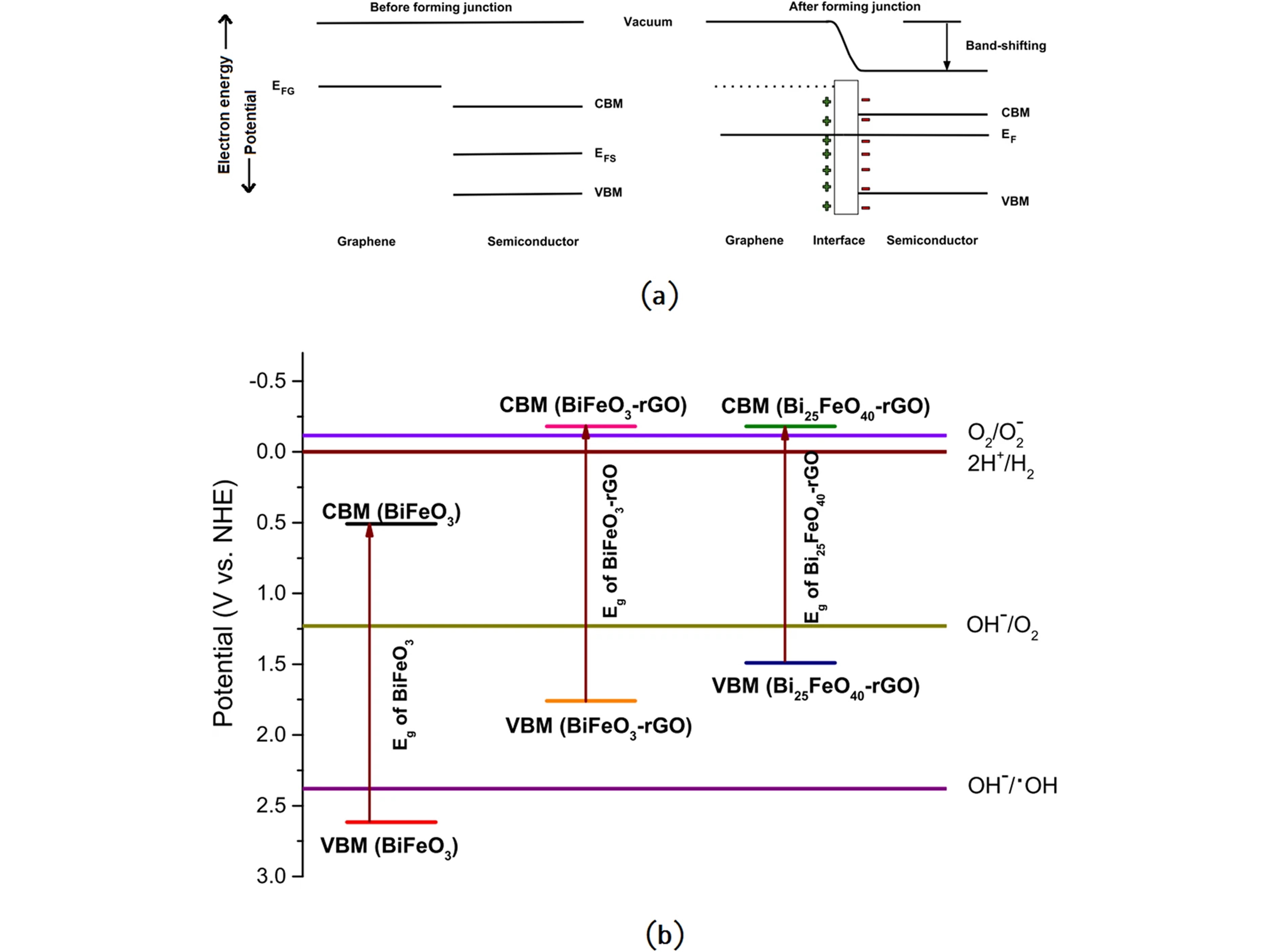}
  \caption{(a) Band-shifting of a nanocomposite while forming a junction with graphene (b) Energy band diagram of BiFeO\textsubscript{3} nanoparticles, BiFeO\textsubscript{3}-rGO, and Bi\textsubscript{25}FeO\textsubscript{40}-rGO nanocomposites.}
\end{figure*}

\subsection{Photocatalytic Mechanism}

The optical band gap is a crucial parameter for gaining an insight into the photocatalytic ability of a material. However, the potentials of the valence band maxima (VBM) and conduction band minima (CBM) also play significant roles in catalyzing the redox reactions in the electrolyte system. Absorption of a photon with energy greater than or equal to the band gap of a material excites an electron from the valence band to its conduction band, i.e., creates an electron-hole pair. The reduction half-reaction commences when the photogenerated electron has a potential that exceeds the reduction potential of the redox couple. Similarly, the photogenerated hole is required to possess a potential that exceeds the redox potential of the oxidation half-reaction. Efficient photocatalytic degradation ability of the nanocomposites can be better understood by analyzing the redox reactions in the electrolyte system that govern the degradation of RhB dye. Photocatalytic degradation of RhB is, in fact, a collection of specific redox reactions. Among these redox reactions, two are particularly important \cite{refXXX}. First, the photogenerated electron can react with the surface adsorbed O$_2$ (redox potential: -0.16 V vs. NHE) to form O$_2^{-}$, which will further react with RhB to cause degradation \cite{refXXX}. Second, the photogenerated holes can react with the OH$^{-}$ ionized from the water molecules to produce $\cdot OH$ (redox potential: 2.38 V vs. NHE). This $\cdot OH$ can further oxidize the RhB molecules. Therefore, a photocatalyst needs to possess a CBM \textless -0.16 V to drive the first reaction and VBM \textgreater 2.38 V to drive the second one efficiently.

\tab These prerequisites necessitate an investigation of the potentials of the VBM and CBM of the synthesized photocatalysts. The Fermi energy of the pure BiFeO$_3$ and Bi$_{25}$FeO$_{40}$ can be calculated theoretically from the equation, E$_F$ = $\chi$ - E$_C$ \cite{refNeth,refBut}. Here, $\chi$ is the absolute electronegativity of the semiconductor which can be found from the geometric mean of the absolute electronegativity of the constituent elements, and E$_C$ is the energy of a free electron in hydrogen scale which is approximately 4.5 eV. While usually, Fermi energy lies midway between the CBM and VBM of a semiconductor, first-principles studies of several sillenites show that the energy of their VBM coincides with the Fermi energy \cite{refACDB}. Hence, the energies of CBM and VBM of BiFeO$_3$ were obtained by simply adding and subtracting 0.5E$_g$ to its Fermi energy whereas for Bi$_{25}$FeO$_{40}$, VBM was considered equal to the Fermi energy and CBM was determined by adding E$_g$ to the VBM. \{CBM, VBM\} of BiFeO$_3$ and Bi$_{25}$FeO$_{40}$ has been determined to be \{0.51 V, 2.61 V\} and \{-0.06 V, 1.74 V\} respectively. Therefore, pure BiFeO$_3$ can perform only the second redox reaction of RhB degradation since it has VBM \textgreater 2.38 V and CBM \textgreater -0.16 V. Infusion of rGO influences the valence and conduction band structure of the pure BiFeO$_3$ and Bi$_{25}$FeO$_{40}$ nanoparticles. Introduction of C 2p orbital of rGO causes both valence and conduction band potentials to raise upward as shown in a previous first-principle investigation on BaTiO$_3$-rGO composite \cite{refADBC}. We propose that similar interactions may have taken place in the BiFeO$_3$-rGO and Bi$_{25}$FeO$_{40}$-rGO nanocomposites. Furthermore, this shift in the band positions can be viewed from a different perspective. Fig. 6(a) shows the formation of a heterojunction between graphene and a semiconductor where Fermi energies of graphene and the semiconductor have been denoted as E$_{FG}$ and E$_{FS}$ respectively. If E$_{FG}$ \textgreater E$_{FS}$, electrons will be transferred to the semiconductor from graphene during the formation of the heterojunction until the system reaches the equilibrium Fermi energy, E$_{F}$ \cite{refADCB}. This electron transfer at the heterointerface will give rise to the formation of a surface dipole \cite{refXOXO,refXOXO2,refrev12} as shown in Fig. 6(a). For a bulk semiconductor, the electric field that arises from the surface dipole will cause band bending that will uniformly change the potentials of vacuum, CBM, and VBM. However, for a nanostructured semiconductor, nearly flatband conditions are observed at the heterojunction since its dimensions are not large enough to change the potentials of CBM and VBM \cite{refXOXO3,refXOXO4}. Consequently, the electric field brings the vacuum level closer to both CBM and VBM. This shift in the potential of vacuum level will reduce the CBM of the nanostructured semiconductor as shown in Fig. 6(a). Since Fermi energy of pristine graphene (0 V vs. NHE) \cite{refACBD,refzyz} is greater than the Fermi energies of both BiFeO$_3$ and Bi$_{25}$FeO$_{40}$, the model described above predicts that upward band-shifting should take place during the formation of both nanocomposites. This upward shift makes the VBM \textless 2.38 V for both nanocomposites, rendering them unable to perform the second redox reaction of RhB degradation using the photogenerated holes. Therefore, the high photocatalytic degradation performance by the nanocomposites should be attributed to the first reaction. Due to the upward band-shifting, CBM potentials of both nanocomposites become sufficiently negative (CBM \textless -0.16 V) which drives the photogenerated electrons in performing the first redox reaction of RhB degradation. Fig. 6(b) consolidates the estimated CBM and VBM potentials of all the synthesized nanomaterials as well as the redox potentials of different redox half-reactions of our interest. Notably, negative CBM potential is one of the most important conditions for efficient photocatalytic hydrogen production via water-splitting. As observed in Fig. 5, our experimental results from photocatalytic hydrogen production test also support the prediction made by our model as the nanocomposites demonstrate considerably higher hydrogen production compared to that of pure BiFeO$_3$ nanoparticles.

\tab Since pure BiFeO$_3$ has a positive CBM potential, it is intuitively confusing to notice that it can also produce hydrogen by water-splitting. A positive CBM potential indicates that only the states at the minima of the conduction band have positive potential. However, there should be several other allowed states in the conduction band with negative potential. When photons with appropriate energy are incident on a material, the electrons of valence band will gain enough energy to reach those states with negative potential and participate in the production of hydrogen. However, these photons, of course, must have energies that are higher than the band gap energy. Therefore, these energies may not lie in the visible range of the solar spectrum. Since the intensity of the incident photons of sunlight decreases considerably with the increase in photon energy beyond visible range, the number of available photons for such transitions is considerably smaller. Besides, these allowed transitions may even be indirect and therefore have smaller transition probability than the near band gap transitions. Therefore, while pure BiFeO$_3$ and several other semiconductors with positive CBM potential may be able to produce hydrogen via water-splitting, their efficiency will be significantly smaller than that of the semiconductors with negative CBM potential. This hypothesis is also evident when we compare the results of photocatalytic degradation of RhB and photocatalytic hydrogen generation. While both BiFeO$_3$-rGO and Bi$_{25}$FeO$_{40}$-rGO nanocomposites exhibit a similar trend of photocatalytic activities during both experiments, BiFeO$_3$ shows a comparatively poorer performance in hydrogen production than that of RhB degradation. This inconsistent performance of BiFeO$_3$ can be better understood if we take a keen look at its VBM potential(2.61 V). Since BiFeO$_3$ has a VBM potential that exceeds 2.38 V, the holes generated by incident photons with energy greater than its band gap of 2.1 eV are sufficient for photodegradation of RhB. However, as mentioned above, photons with much higher energy than 2.1 eV are necessary to commence the production of hydrogen. This inherent shortcoming of BiFeO$_3$ nanoparticles eventually reduces their efficiency in photocatalytic hydrogen production.

\tab Due to the high electronegativity of oxygen atoms, many metal oxides with narrow band gaps such as $\alpha$-Fe$_2$O$_3$, BiVO$_4$, WO$_3$, etc. have positive CBM potentials that suppress their ability in water-splitting. Our proposed model predicts that formation of nanocomposites with rGO may shift their CBM potentials to negative values and trigger the suppressed photocatalytic hydrogen production ability of these metal oxides. This model unveils a novel mechanism behind the superior photocatalytic abilities of semiconductor-rGO nanocomposites.

\tab Effective suppression of recombination processes is another important characteristic of a good photocatalyst. BiFeO$_3$ is a ferroelectric material that possesses a spontaneous inherent electric field. This electric field can spatially separate the photogenerated electrons and holes to reduce the probability of recombination \cite{refE}. A previous first-principles investigation revealed that the recombination processes in a sillenite are suppressed due to an intraband transition of the photogenerated hole \cite{refACDB}. We believe that this suppression of recombination processes would also apply to Bi$_{25}$FeO$_{40}$. Besides, the formation of a composite with rGO causes the photogenerated electrons to move to the rGO sheet where the superior electron mobility of the restored honeycomb structure can help the electron transport to the redox couples \cite{refXX,refYY,refZZ,ref42,refrev12}. As shown in Fig. S3, steady-state photoluminescence(PL) spectra of the photocatalyst samples justify our hypothesis as the nanocomposites show considerably smaller PL peak intensities compared to that of pure BiFeO$_3$ nanoparticles. Smaller PL peak intensity implies that a smaller number of photogenerated electrons and holes are radiatively recombined. Therefore, we believe that the photogenerated electron-hole pairs are more effectively separated when a composite with rGO is formed.

\section{Conclusions}
We have demonstrated a low temperature versatile hydrothermal technique for synthesizing highly efficient visible light photocatalysts of bismuth ferrite-rGO nanocomposite family whose phase can be controlled by tuning the synthesis temperature. Merging rGO sheets with each of the pristine BiFeO$_3$ and Bi$_{25}$FeO$_{40}$ nanoparticles has resulted in better morphology, superior optical absorption, and favorable band edge shifts. The improvements associated with inclusion of rGO have also been reflected on the photocatalytic ability of the nanocomposites as they demonstrated considerably better degradation of RhB dye as well as greater hydrogen production over both pure BiFeO$_3$ nanoparticles and a standard photocatalyst, Degussa P25 titania nanoparticles. Furthermore, Bi$_{25}$FeO$_{40}$-rGO nanocomposite has shown 3.5 times greater photocatalytic hydrogen production compared to that of Degussa P25 titania nanoparticles. Moreover, a theoretical model has been proposed to explain the band-shifting phenomenon which may prove to be significant in understanding the superior photocatalytic performance of different metal oxide-rGO nanocomposites.

\section{Acknowledgments}

This work was financially supported by Ministry of Education, Government of Bangladesh (Grant No. PS 14267).

\section{Appendix A. Supplementary data} 
 Supplementary data to this article can be found online.

\section{Data availability}
The raw and processed data required to reproduce these findings cannot be shared at this time due to technical or time limitations.


\bibliography{PCCP} 
\bibliographystyle{rsc} 


\end{document}